\documentclass[openacc]{rsproca_new}%%%%where rsproca is the template name

\usepackage{multicol}
\usepackage{graphicx} 
\usepackage{authblk}
\jname{rsif}
\Journal{J R Soc Interface}

\begin{document}
%%%% Article title to be placed here
\title{Collective rotational motion of freely-expanding T84 epithelial cell colonies}

\author{%%%% Author details
Flora Ascione$^{1}$, Sergio Caserta$^{1,2}$, Speranza Esposito$^{1,3}$, Valeria Rachela Villella$^{1,3}$
, Luigi Maiuri \thanks{Deceased.}$^{3}$, Mehrana R. Nejad$^{4}$, Amin Doostmohammadi$^{5}$, Julia M. Yeomans $^{4}$, Stefano Guido $^{1,2}$
}

%%%%%%%%% Insert author address here
\address{$^{1}$Dipartimento di Ingegneria Chimica, dei Materiali e della Produzione Industriale (DICMAPI), Università di Napoli Federico II. P.le Tecchio 80, 80125 Napoli, Italy\\
$^{2}$CEINGE Biotecnologie Avanzate, Via Sergio Pansini, 5, 80131 Naples, Italy\\
$^{3}$European Institute for Research in Cystic Fibrosis, San Raffaele Scientific Institute, Milan, Italy
\\
$^{4}$ The Rudolf Peierls Centre for Theoretical Physics, Department of Physics, University of Oxford, Parks Road, Oxford OX1 3PU, UK
\\
$^{5}$ The Niels Bohr Institute, University of Copenhagen, Copenhagen, Denmark
}

%%%% Subject entries to be placed here %%%%
\subject{biophysics}

%%%% Keyword entries to be placed here %%%%

%%%% Insert corresponding author and its email address}
\corres{Author for correspondence:\\
\email{sergio.caserta@unina.it}\\
\email{julia.yeomans@physics.ox.ac.uk}}

%%%% Abstract text to be placed here %%%%%%%%%%%%
\begin{abstract}
Coordinated rotational motion is an intriguing, yet still elusive mode of collective cell migration, which is relevant in pathological and morphogenetic processes. Most of the studies on this topic have been carried out on epithelial cells plated on micropatterned substrates, where cell motion is confined in regions of well-defined shapes coated with extracellular matrix adhesive proteins. The driver of collective rotation in such conditions has not been clearly elucidated, although it has been speculated that spatial confinement can play an essential role in triggering cell rotation. Here, we study the growth of epithelial cell colonies freely expanding (i.e., with no physical constraints) on the surface of cell culture plates and focus on collective cell rotation in such conditions, a case which has received scarce attention in the literature. One of the main findings of our work is that coordinated cell rotation spontaneously occurs in cell clusters in the free growth regime, thus implying that cell confinement is not necessary to elicit collective rotation as previously suggested. The extent of collective rotation was size and shape dependent: a highly coordinated disk-like rotation was found in small cell clusters with a round shape, while collective rotation was suppressed in large irregular cell clusters generated by merging of different clusters in the course of their growth. The angular motion was persistent in the same direction, although clockwise and anticlockwise rotations were equally likely to occur among different cell clusters. Radial cell velocity was quite low as compared to the angular velocity, in agreement with the free expansion regime where cluster growth is essentially governed by cell proliferation. A clear difference in morphology was observed between cells at the periphery and the ones in the core of the clusters, the former being more elongated and spread out as compared to the latter. Overall, our results provide the first quantitative and systematic evidence that coordinated cell rotation does not require a spatial confinement and occurs spontaneously in freely expanding epithelial cell colonies, possibly as a mechanism for the system. 
\end{abstract}
%%%%%%%%%%%%%%%%%%%%%%%%%%%

%%%%%%%%%% Insert the texts which can accomdate on firstpage in the tag "fmtext" %%%%%

%%%%%%%%%%%%%%% End of first page %%%%%%%%%%%%%%%%%%%%%

\maketitle

\begin{multicols}{2}

\section{Introduction}
 The cooperative movement of cell groups, sheets or strands, which is referred to as collective cell migration \cite{one}, plays a key role in several physiological as well as pathological processes including morphogenesis, tissue repair, immune response, and cancer progression \cite{vedel2013migration,friedl1995migration,friedl2009collective,C4IB00115J}. In tumor invasion, for example, collective cell movement allows malignant tumor cells to escape the primary tumor and invade surrounding tissues \cite{rorth2009collective,friedl2012classifying}.
Similarly to single cell migration, collective cell movement is mainly driven by actomyosin polymerization and contractility coupled to cell polarity \cite{ilina2009mechanisms}, but occurs under additional constraints, determined by cell-cell junctions and other close cell interactions \cite{ilina2009mechanisms,rorth2007collective}. The latter include several processes that have been proposed to affect collective migration, such as direct cell-cell chemical signaling, physical interactions underlying the mechanical integrity of clusters, the coordinated polarization of leader cells on cluster edges possibly guiding the behavior of follower cells, and the secondary remodeling of the extracellular matrix along the migration track \cite{ilina2009mechanisms}. Traction force mapping shows long-range force transmission within sheets or clusters in a cooperative way: each cell, at the leading edge as well as inside, takes part in a global “tug-of-war” through cell-cell junctions that maintains the collective structure in a global state of tensile stress \cite{trepat2009physical,ladoux2009cells,ladoux2012physically}. Physical signals from the substrata, such as local rigidity \cite{beaune2014cells}, have also an effect on cell migration. 

The complex interplay between different processes over a wide range of spatiotemporal scales makes the interpretation of cooperative cell behavior a challenging task. Some important insights have been provided by suggesting several analogies with other phenomena, such as the flow of multiphase fluids (e.g., the spreading and coalescence of droplets \cite{beaune2014cells}), the mechanical behavior of liquid crystals \cite{gompper20202020,mueller2019emergence}, and the rheology of the glassy state \cite{angelini2010cell}. These analogies are based on the fact that cell-cell junctions are fluid in the sense that they can be mutually displaced, thus allowing cell motion even in dense aggregates. Hence, one can describe cell motility in analogy with temperature-driven molecular motion and introduce a diffusion coefficient to model cell migration against a cell density gradient. However, two important differences arise with respect to the case of ordinary fluids: cells are able to propel themselves in a given direction and can proliferate, which are two typical attributes of living active matter \cite{zorn2015phenomenological}. Therefore, collective cell migration can be considered as the result of three main effects: cell proliferation, diffusive migration and directional motion. The latter can be elicited by some substrate anisotropy, such as chemotaxis and contact guidance. \\

A mode of collective cell migration that has attracted much interest in the literature is the coordinated rotation of cells, also referred to as coherent angular motion (CAM) \cite{tanner2012coherent}. Such rotational motions have been observed in human breast epithelial cells cultured in laminin-rich gels in vitro and interpreted as an essential mechanism for the formation of acini, polarized spherical structures with basolateral and apical membrane regions around a central lumen found in mammary glands \cite{tanner2012coherent}. Lumen formation in vitro by MDCK (Madin-Darby canine kidney) epithelial cells aggregates has also been associated with circular cell motility \cite{ferrari2008rock}, with a rate decreasing with increasing cell number, suggesting a transition to epithelial polarization during aggregate development \cite{marmaras2010mathematical}. In vivo, collectively rotating cell structures have been observed during morphogenesis, such as in the development of the primitive streak in gastrulating chick embryos \cite{vasiev2010modeling}. 

Most recent studies on collective cell migration, including CAM, have been carried out by constraining cell motion inside confined regions obtained by micropatterning islands coated with extracellular matrix adhesive proteins (e.g., see \cite{huang2005symmetry,vedula2012emerging,deforet2014emergence,marel2014flow,zorn2015phenomenological,peyret2019sustained}). As opposed to the classical wound healing assay, where the time taken by the cells to fill a gap created by scratching a confluent cell monolayer is measured as an index of collective cell migration \cite{caserta2013methodology}, the technique of micropatterning allows a better control of domain geometry both in terms of size and shape (e.g., circular and rectangular regions of different size can be created). In addition, cell motion inside micropatterned surfaces can be investigated by several techniques, such as time-lapse microscopy, epifluorescence and confocal imaging, Particle Image Velocimetry and image analysis, enabling to obtain spatiotemporal maps of cell position, orientation and shape with subcellular resolution. This experimental characterization can be compared with theoretical modeling of collective cell migration. The latter has been based on several approaches, including continuum modeling by the Fisher-Kolmogorov equation \cite{marel2014flow,ascione2017wound}, phenomenological migration motives \cite{kim2013propulsion,trepat2011plithotaxis} (e.g., plithotaxis and kenotaxis), and the already mentioned analogies with glassy dynamics \cite{angelini2010cell}, droplet-like spreading on rigid surfaces \cite{beaune2014cells,geo}, and active nematic liquid crystals \cite{doostmohammadi2015celebrating,prost2015active,doostmohammadi2016defect}.

Under confined conditions in circular patterns, MDCK cells are found to exhibit solid-body behavior of synchronized collective rotation when they reach confluency \cite{doxzen2013guidance,doostmohammadi2019coherent}. It has been speculated that such behavior is initiated by cells at the border of the circular regions, which are guided by the edge and tend to transmit their directional motion to inner cells through cell-cell contacts \cite{doxzen2013guidance}. Indeed, by downregulating intercellular adhesion the synchronized collective cell rotation is reduced. Cell-cell contacts provide directional guidance to neighbor cells, so that the ring orientation at the periphery of the circular patterns is propagated to inner cells. This mechanical coupling is altered by the presence of a cell in the center of the circular pattern due to the lack of a stable axis of internal polarization, which can explain the discontinuity in the persistence time of CAM between 4 and 5 cells (when a central cell is observed in the circular pattern) \cite{segerer2015emergence}. Another factor playing a key role in CAM is cell division, which has been shown to induce extensile forces and turbulent-like velocity fields in confluent cell monolayers \cite{doostmohammadi2015celebrating,siedlik2017cell}. Cell division is also associated with a switching of the direction of collective cell rotation \cite{siedlik2017cell}. Blocking cell proliferation with mitomycin-C impairs CAM and the switches of the direction of rotation. Division of a cell located near the periphery is more effective in inducing the onset of CAM as compared to the division of inner cells \cite{siedlik2017cell}. 

Although much progress has been made in the understanding of collective cell rotation, there are still several aspects to be fully elucidated. In particular, while some consensus has been reached on the role of confinement in inducing CAM, the same does not apply to the driver of such cell collective behavior, for which an intriguing explanation could be spontaneous symmetry breaking. In principle, some further insight on collective cell rotation could be obtained by looking at unconstrained expansion of cell colonies, which has been less studied in the literature as compared to the case of confined conditions. In a work addressing the topic of freely expanding epithelial MDCK cells \cite{puliafito2012collective}, it has been shown that in the first 5-6 days the area of a colony grows according to a simple exponential law while cell density remains constant. Furthermore, cells move outwards on the average, with nonuniform velocity at the periphery and finger-like protrusions. At a critical value of area, the outwards expansion of the colony cannot keep pace with cell proliferation and cell density starts increasing. At this point, a transition to an epithelial morphology is observed, where cell height is increased, local ordering appears in the colony and cell proliferation and motion are strongly inhibited \cite{puliafito2012collective}. Overall, these results provide a quantitative characterization of collective cell migration and growth in freely expanding cell colonies, but do not address coordinated cell rotation. 

Indeed, data on coordinated cell rotation in this free growth regime are still lacking in the literature and the main objective of our work is to study such phenomenology. 
Here, we focus on collective rotation in freely-expanding epithelial colonies of the T84 intestinal cell line, where, at variance with micropatterned substrata, cell clusters can grow with no geometrical constraints. The T84 cells are seeded in multiwell culture plates and the evolution of the growing cell clusters is followed up to confluency by time-lapse video microscopy and image analysis. The latter allows to track the position of individual cells as a function of time and to investigate their angular motion inside the clusters. The results of this work are relevant in several fields, from tissue morphogenesis to the repair of injuries of the epithelial layer lining the intestinal lumen.

\section{Materials and Methods}
\subsection{Cell cultures}
    Human colon adenocarcinoma T84 cells were cultured in Dulbecco’s Modified Eagle Medium (DMEM) F12 supplemented with $10\%$ (v/v) Fetal Bovine Serum (FBS) and antibiotics (50 units/mL penicillin and 50 µg/mL streptomycin) and maintained in a humidiﬁed incubator at 37 °C under an atmosphere of $5\%$ $CO_2$ in air.

\subsection{Time Lapse Microscopy}
T84 cells were plated on 24 multi-well culture dishes at varying cell density. The multi-well plate was placed on the stage of an inverted microscope (Zeiss Axiovert 200) equipped with motorized sample positioning and focusing (Marzhauser). The microscope was enclosed in a plastic cage to control environmental conditions (temperature: 37°C, $CO_2$  concentration: $5\%$, humidity level: close to saturation) \cite{caserta2013methodology,silano2012early}. Live cell phase contrast imaging was performed by a high resolution monochromatic CCD video camera (Hamamatsu Orca AG). Microscope operations were controlled by a time-lapse software allowing to select multiple fields of view and the time interval between consecutive image acquisitions during the experiment \cite{ascione2014investigation,vasaturo2012novel}. To follow T84 collective behavior, images were iteratively acquired using a 5x objective at several locations within the culture dish, with an image acquisition frequency of 2 h; the overall experimental length was 14 days. The cells were rinsed with fresh culture medium every two days, without removing the culture dish from the microscope stage. To capture the dynamic behavior of individual cells within the aggregates, images were iteratively acquired using a 20x objective. Acquisition frequency was 10 min, the experiment length was 70 h.

%\subsection{Confocal imaging}
%Confocal images were acquired by using a confocal laser scanning microscope (LSM 5 Pascal, Zeiss) equipped with a helium/neon laser (LASOS Lasertechnik GmbH, LGK SAN7460A). Observations were carried out with a 40x objective (Zeiss) with a numerical aperture of 0.9.
\begin{figure*}[t]
    \centering
    \includegraphics[width=0.59\linewidth]{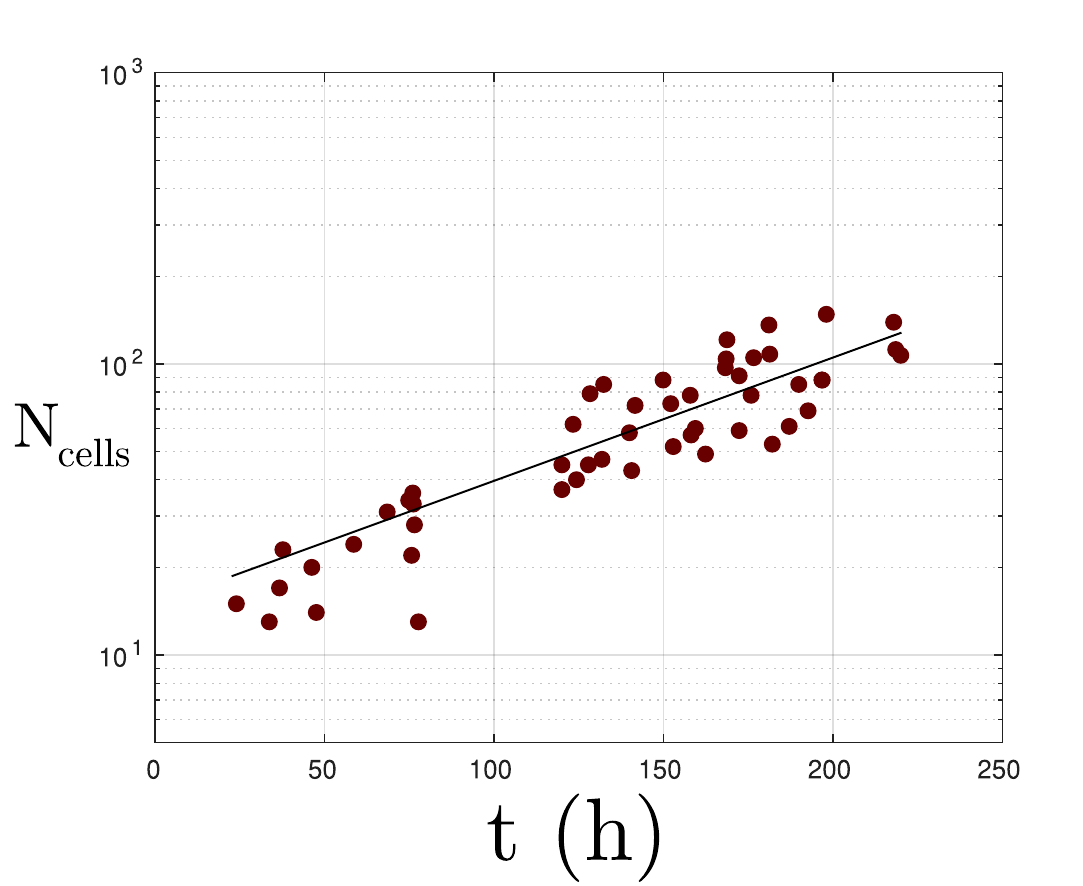}
      \caption{Dynamic evolution of the number of cells within the clusters. The number of cells ($N_{cells}$) is plotted as a function of time. An exponential fit is shown as a continuous line.}\label{fig:1}
\end{figure*}
\subsection{Image analysis}
The number of cells ($N_{cells}$) within each cell cluster was determined by using the Image Pro Plus analysis software, which allows to manually count the cells in a selected region. The same software was used to measure the area of the cell clusters ($A_{cluster}$ ) by manually tracing their contour in the image overlay. The average area of individual cells inside a cluster ($A_{cell}$) was determined as the ratio between the area of the cluster ($A_{cluster}$ ), and the number ($N_{cells}$) of enclosed cells.

By using semiautomatic routines (ImageJ), the X and Y coordinates of the centroid of individual cells were manually measured at each time step, and the corresponding cell trajectories were reconstructed. In order to quantify cell movement in a cluster, the radial ($V_{\rho}$), and angular ($\omega$) cell velocity were calculated from the net cell displacement with respect to the cluster centroid, along the radial and the angular direction, respectively, over a time interval of 2 h. The data were then averaged over the entire cell population. The angles were measured counterclockwise and cell tracking was done for 12 h.

\subsection{Equations of motion for the continuum model}
Active nematic descriptions have been very successful in describing the dynamics of cell monolayers \cite{saw2017topological,saw2018biological,guillamat2022integer}. 
The fundamental continuum equations that describe wet active nematics (active nematohydrodynamic equations) are coupled equations for the evolution of the cell concentration $\phi$, nematic tensor, $\textbf{Q}=S(\textbf{n}\textbf{n}-\textbf{I}/2)$ in two dimensions, and the associated incompressible fluid velocity, $\mathbf{u}$. They read
\begin{align}
   &\partial_t \textbf{Q} + \textbf{u} \cdot \boldsymbol{\nabla} \textbf{Q}- \boldsymbol{\mathcal{W}}= \gamma \: \textbf{H},\label{n:qevoln}\\
    &\rho \left(\partial_t + \textbf{u}\cdot \boldsymbol{\nabla}\right) \textbf{u}= \boldsymbol{\nabla} \cdot \boldsymbol{\Pi} , \quad \boldsymbol{\nabla}\cdot \textbf{u}=0,\label{eqn:nsbb}\\
    & \partial_t \phi + \textbf{u} \cdot \boldsymbol{\nabla} \phi =\Gamma_{\phi} \nabla^2 \mu,\label{eqn:nscf}\\
     & \mu= \frac{\partial f}{\partial \phi}-\boldsymbol{\nabla} \cdot (\frac{\partial f}{\partial \boldsymbol{\nabla} \phi}),\label{eqn:nsd}\\
     & f=  -\frac{C}{2}(1+Q_{ij} Q_{ij}/2)^2+\frac{B}{2}\phi^2(1-\phi^2) \\
     & +\frac{K_{\phi}}{2} \nabla_m \phi \nabla_m \phi +\frac{K}{2} \nabla_m Q_{ij} \nabla_m Q_{ij}.\label{eqn:nsdf}
\end{align}
In the definition of the nematic tensor the director field $\textbf{n}$ represents the orientation of the nematic alignment and the magnitude of the nematic order is denoted by $S$. In the evolution of the $\textbf{Q}$ tensor, Eq.~(\ref{n:qevoln}), $\gamma$ is the rotational diffusivity and the molecular field, $\mathbf{H}= -\partial f/\partial \textbf{Q}+ \boldsymbol{\nabla} \cdot (\partial f/\partial \boldsymbol{\nabla} \textbf{Q})$, drives the nematic tensor towards the minimum of a free energy density $f$. The generalized advection term  
 \begin{align}\label{elastic}
    \boldsymbol{\mathcal{W}}=&(\lambda \textbf{E}+\boldsymbol{\Omega})\cdot(\textbf{Q}+\frac{\textbf{I}}{2})+(\textbf{Q}+\frac{\textbf{I}}{2})\cdot(\lambda \textbf{E}-\boldsymbol{\Omega}) 
    \\ &-\lambda(\textbf{Q}+\textbf{I})Tr(\textbf{Q} \cdot \textbf{E})
\end{align} 
 models the response of the nematic field to the strain rate $\textbf{E}$ and vorticity $\boldsymbol{\Omega}$, where $\lambda$ is the flow-aligning parameter. 
\begin{figure*}[t] % only one legend, bigger tick markers, make each graph higher
    \centering
    \includegraphics[width=0.59\textwidth]{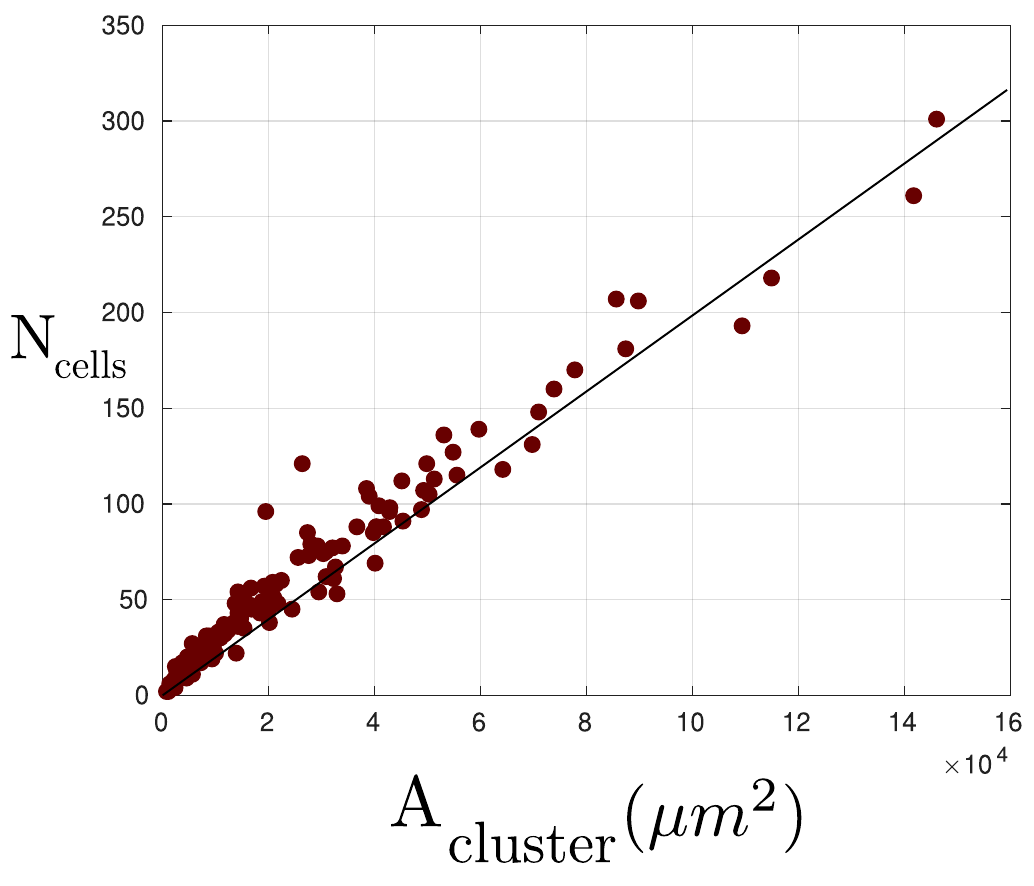}
    \caption{The number of cells ($N_{cells}$) within the clusters is plotted as a function of the cluster area ($A_{cluster}$). The continuous line is a linear fit to the data.}\label{fig:2}
\end{figure*}
In the Navier-Stokes equation (\ref{eqn:nsbb}), $\rho$ is the density of the suspension, and the stress tensor, $\boldsymbol{\Pi}$, includes viscous, elastic and active contributions. The viscous stress, $\boldsymbol{\Pi}^{v}=2\eta \textbf{E}$, where $\eta$ is the viscosity, and the elastic stress,
\begin{align}\label{ela}
\Pi^{p}_{ij} =& -P \delta_{ij} + \lambda (Q_{ij}+\delta_{ij} ) Q_{kl} H_{kl} -\lambda  H_{ik} (Q_{kj}+\frac{\delta_{kj}}{2})\nonumber \\ & -\lambda(Q_{ik}+\frac{\delta_{ik}}{2}) H_{kj}+ Q_{ik} H_{kj}- H_{ik} Q_{kj}\nonumber \\& - K (\partial_i Q_{kl})(\partial_j Q_{kl}),
\end{align} 
where  $P$ is the pressure, are familiar terms that appear in the dynamical equations of passive liquid crystals. 
Coarse-graining the dipolar force fields of the active nematogens leads to an active contribution to the stress that characterises wet active nematics, $\boldsymbol{\Pi}^{\text{a}}=\zeta \textbf{Q}$. In extensile systems ($\zeta<0$), it acts to extend a nematic region along its director whereas in contractile materials ($\zeta>0$) it contracts a nematic region along the director \cite{PhysRevLett.128.048001}.
Eq.~(\ref{eqn:nscf}) describes the evolution of the concentration of the active material, where $\Gamma_{\phi}$ shows how fast $\phi$ responds to gradients in the chemical potential $\mu$.

The first term in the free energy density (Eq.~(\ref{eqn:nsdf})) leads to the isotropic phase so that in our system, any nematic order is caused by activity ~\cite{PhysRevLett.125.218004,santhosh2020activity}. 
The second and third terms lead to the formation of cell clusters (identified by $\phi=1$) in a  cell-free background ($\phi=0$), where $C$, $B$ and $K_{\phi}$ are material parameters. The last term in Eq.~(\ref{eqn:nsdf}) represents the energy cost due to distortions in the nematic field, assuming a single Frank elastic constant $K$. 

The continuum equations of motion are solved using a hybrid lattice Boltzmann and finite difference method \cite{PhysRevE.76.031921,PhysRevLett.124.187801,nejad2020memory}.

\subsection{Simulation parameters}
For the simulations, we used a box of size $300\times300$ in lattice Boltzmann units with periodic boundary conditions, and start the simulations with circular drops of various sizes with a random director (the initial orientation is randomly chosen and is in the interval $[0, 2\pi]$). We used the parameter values $\rho=40$, $\Gamma_{\phi}=0.2$, $\gamma=0.3$, $|\zeta|=0.01$, $|\chi|=0.7$, $K=0.02$, $K_{\phi}=0.1$, $B=0.01$, $C=0.001$. To compare the rotational velocity as a function of area with the experimental data, we matched the initial area and angular velocity in simulations with the initial area and angular velocity in the experiment and rescaled the axis accordingly.

\section{Results and discussion}
Previously some of us have investigated the dynamics of monolayer formation in human colon adenocarcinoma T84 cells in vitro by using time-lapse microscopy \cite{silano2012early}. As summarised in the Introduction, the expansion of an epithelial colony is generally driven by a complex interplay of proliferation \cite{mehes2012collective}, cell growth \cite{puliafito2012collective}, motility \cite{ouaknin2009stochastic} and cluster fusion mechanisms \cite{douezan2012active}. At early times upon seeding T84 cells in culture plates, 2D cell clusters or islands of different size, depending on the initial cell density, were observed. Initially, cell clusters were separate from each other and their size increased in time without apparent mutual interactions. As time went on, however, cell clusters came in contact with each other during their growth and merged together in larger clusters, up to the eventual formation of a cell monolayer spanning the entire available surface of the culture plate. It was also observed that cells at the cluster boundary extended long filopodia reaching out to nearby clusters before their merging. A quantitative characterization of the expansion and growth of T84 cell clusters up to the formation of a continuous monolayer was carried out by measuring the number of cells within the clusters and the cluster size as a function of time. Furthermore, the motility of single cells within the clusters was analyzed as described in the previous section.

\subsection{Cell proliferation analysis}
In Fig.~\ref{fig:1} the number of T84 cells inside a cluster, $N_{cells}$, is plotted as a function of time. It can be noticed that data from different cell clusters are pooled together in the plot of Fig. \ref{fig:1} and, apart from some scatter, they seem to follow the same trend. The latter is well represented by a simple exponential law:
\begin{equation}\label{eq1}
N_{cells} (t)= N_0 t^{t/t_d} 
\end{equation}
where t is the time from cell plating, $N_{cells}$(t) the number of cells at time t, $N_0$  the number of cells at time 0 and $t_d$ the cell duplication time. Eq.~(\ref{eq1}) was fit to the experimental data in Fig. \ref{fig:1}, with $t_d$ as the only adjustable parameter. This approach neglects the contribution of cell death to the cell population balance, assuming that it is not relevant in the initial phase of cell growth. The exponential fitting curve is shown as a continuous line in the graph in Fig.~\ref{fig:1}. Based on the fit, the duplication time of the cells within the clusters was estimated to be 71 h (with $R_2 = 0.71$ and a standard error of estimate of 19 h). The observed trend does not follow the classical logistic cell growth, which is characterized by a sigmoidal shape with a decreased rate when cell density becomes so large that further proliferation is inhibited. The time frame of the experiments is indeed up to the onset of confluency and the experiments are not carried out afterwards (see next subsection), since our focus is to study coordinated cell rotation in isolated cell clusters, i.e., well before confluency.  
\begin{figure*}[t] % only one legend, bigger tick markers, make each graph higher
    \centering
    \includegraphics[width=0.99\textwidth]{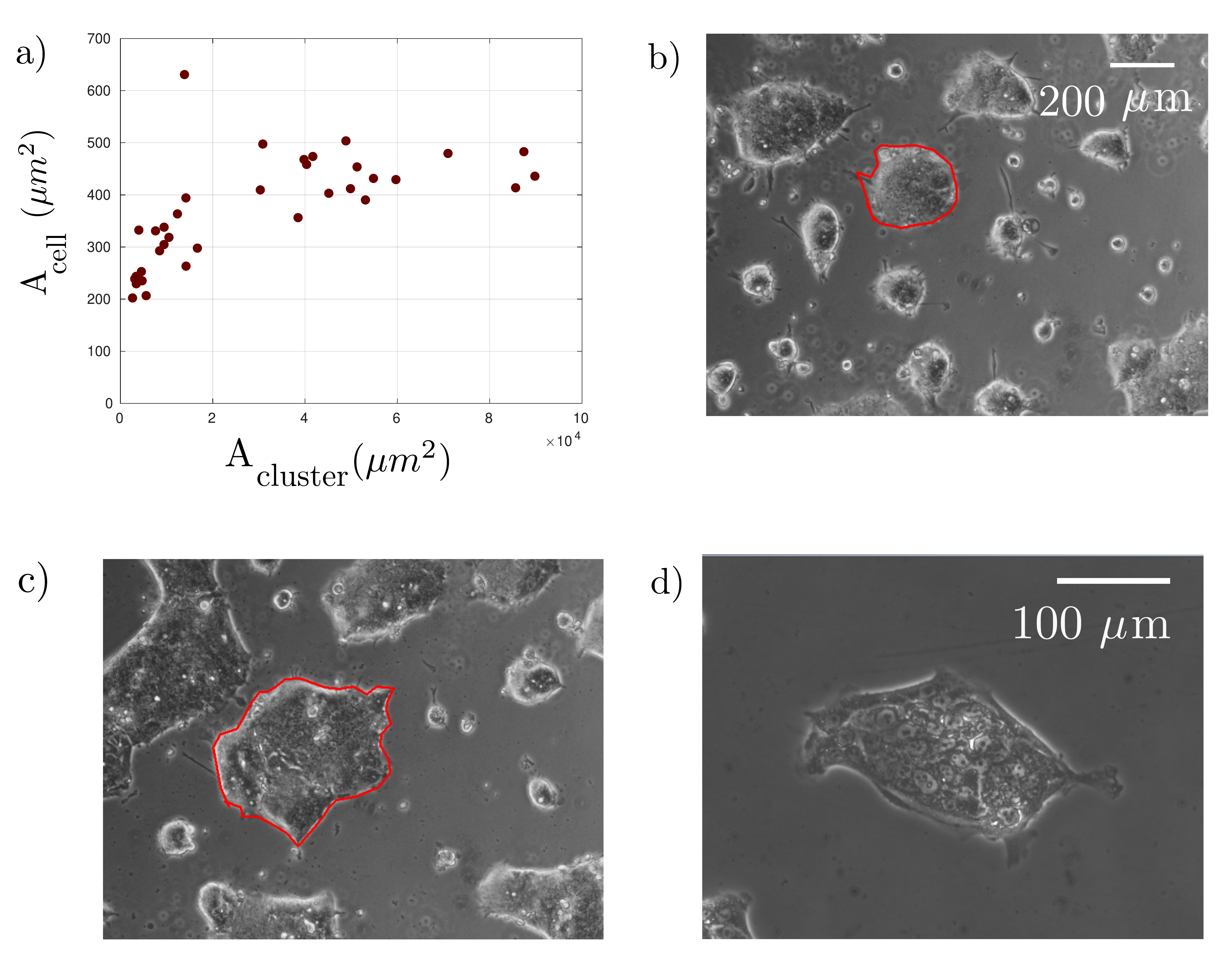}
    \caption{Average cell area is plotted as a function of cluster area (a). Two images of the same field of view at t=0 (b) and t=24 h (c) are shown below the plot. The contour of a cell cluster is highlighted as a red overlay in both images (scale bar = 200 $\mu m$). A larger view of a cell cluster showing the different morphology of cells at the edge and in the core (scale bar = 100 $\mu m$) is presented in (d).}\label{fig:3}
\end{figure*}
\subsection{Growth of cell clusters}
The morphology of cell clusters was studied by manual tracing of their boundaries followed by application of image analysis routines (see Materials and Methods). The resulting cell cluster area, $A_{cluster}$, is correlated to $N_{cells}$, the number of cells within a cluster, in Fig.~\ref{fig:2}, where results corresponding to 51 clusters are presented and the continuous line is a linear fit to the data. The plot in Fig.~\ref{fig:2} shows a direct proportionality between $A_{cluster}$  and $N_{cells}$, thus implying that the average cell density in a cluster is constant in the time frame investigated. This finding is consistent with the so called free expansion regime, where the increase of cluster area is governed by mitosis and each daughter cell takes an area equal to that of the mother cell \cite{puliafito2012collective}. 

\begin{figure*}[t] % only one legend, bigger tick markers, make each graph higher
    \centering
    \includegraphics[width=0.99\textwidth]{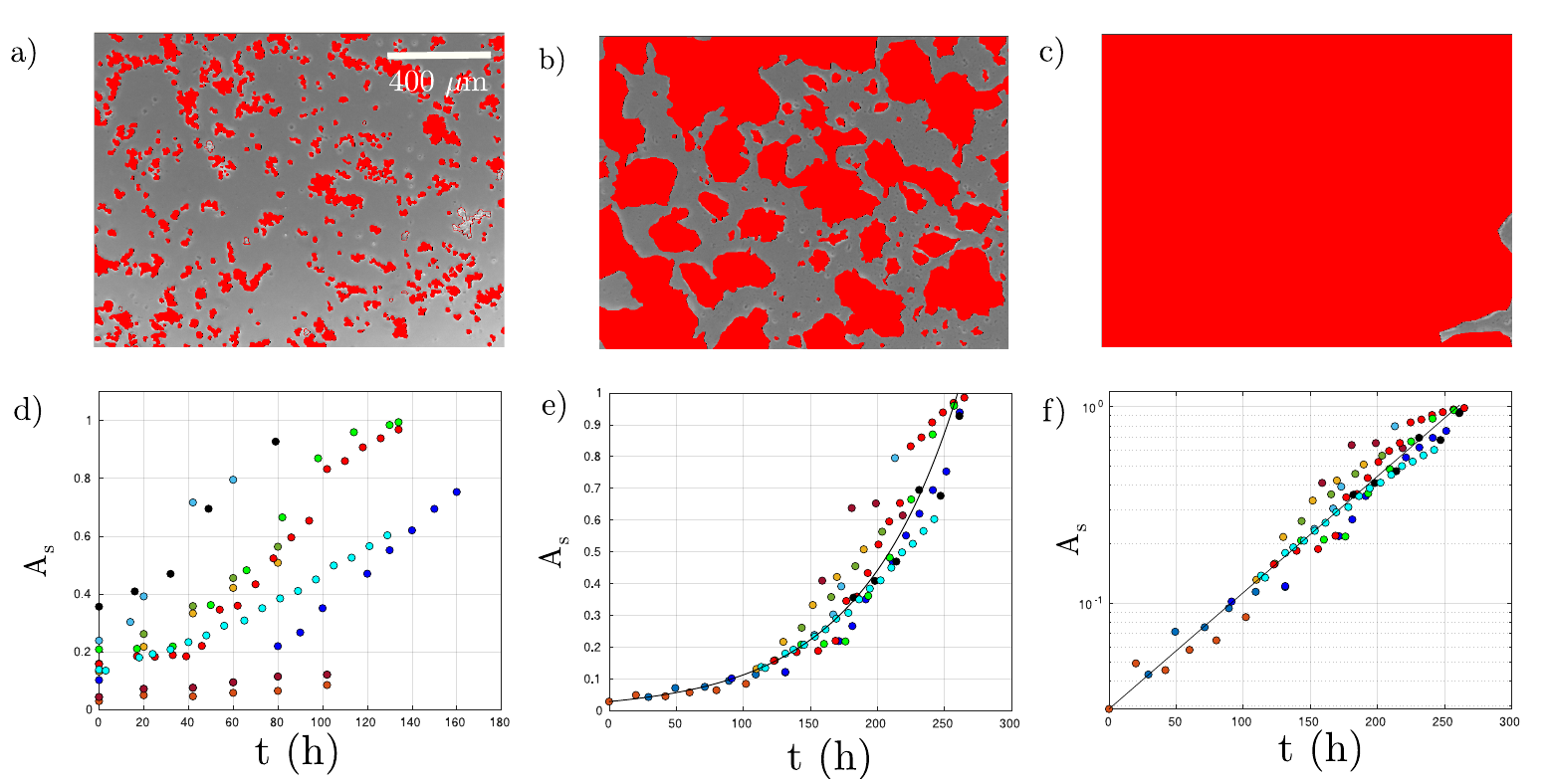}
    \caption{a)-c) Images of monolayer spreading at time t=0 (a), t=80 h (b) and t=160 h (c), corresponding to $A_s=0.10$, $A_s=0.50$ and $A_s=0.99$ respectively, are shown, with the area occupied by the cells highlighted in red. This is different from the scale shown on the image: Scale bar is equal to 400 $\mu m$. Dynamics of monolayer formation for different initial cell densities. Cell confluence ($A_s$) is plotted as a function of time for each cell sample under investigation. On the left (d) raw data from each sample are reported, on the right (e) the same data are shifted in time to generate a master curve, that can be used to estimate the characteristic time of the process. }\label{fig:4}
\end{figure*}

As for $N_{cells}$, the cluster area also grows exponentially (data not shown for the sake of brevity). The area exponential growth cannot continue indefinitely, since it would require an unbound exponential increase of cell velocity. At some point, cell velocity cannot keep up with area growth due to cell proliferation, and cell density starts increasing, which marks the onset of the proliferation inhibition mentioned in the previous subsection. Following \cite{puliafito2012collective}, the critical cluster area $A_c$ at the transition between these two regimes can be estimated by the equation
$v_{\rho}=(A_c/(4 \pi \tau^2))^{1/2}$, where $v_{\rho}$ is the radial cell velocity at the boundary and $\tau$ is the characteristic time of the exponential area growth. From this equation and the values $\tau=t_d=71 h$ (see previous subsection) and $ v_{\rho} = 0.04 \mu m/min = 2.4 \mu m/h$ (see the subsection on cell motility), one obtains $A_c \sim 4 \times 10^5 \mu m^2$ , which is above the largest value ($1.5 \times 10^5 \mu m^2$) of cluster area in Fig.~\ref{fig:2}, thus confirming that the range explored in our experiments is within the free expansion regime.

A further analysis of cell density was performed by dividing cluster area by the corresponding number of cells and plotting the resulting average cell area $A_{cell}$ as a function of cluster area $A_{cluster}$, as shown in Fig.~\ref{fig:3}.

It can be noticed that $A_{cell}$, which is the reciprocal of cell density, is constant with cluster area apart from the initial part of the plot, where the data of cell number in Fig.~\ref{fig:2} are slightly above the linear fit. This trend can be explained by the spreading of cells at the boundary, as discussed in the following.
The time evolution of clusters can be monitored by time-lapse video microscopy of the growing colony, as illustrated in the images at the bottom of Fig.~\ref{fig:3}. The image in Fig.~\ref{fig:3}b, corresponding to the beginning of the time-lapse experiment, is characterized by smaller clusters with bright boundaries in phase contrast, indicating thicker cells.

The clusters in Fig.~\ref{fig:3}b, corresponding to a time of 24 hours after the beginning of the time-lapse experiment, are less bright at the boundary due to the spreading of the cells therein. Several cell protrusions, such as filopodia and lamellipodia, between nearby clusters can be observed in both images of Fig.~\ref{fig:3}b and c, and provide a further argument in favor of the special role played by cells at the boundary. The image in Fig.~\ref{fig:3}d shows a larger view of a single cluster and it can be observed that cells at the boundary are more elongated and flattened as compared to cells in the middle. Such different morphology has been linked to the hypothesis that cells at the periphery can act as leaders dragging along the cells which are located inside the clusters, although this view is still a matter of debate \cite{qin2021roles}. Another possible reason for the difference of morphology is “active anchoring” - in an active extensile system the cells would tend to elongate parallel to the boundary \cite{PhysRevLett.113.248303}.

\begin{figure*}[t] % only one legend, bigger tick markers, make each graph higher
    \centering
    \includegraphics[width=0.99\textwidth]{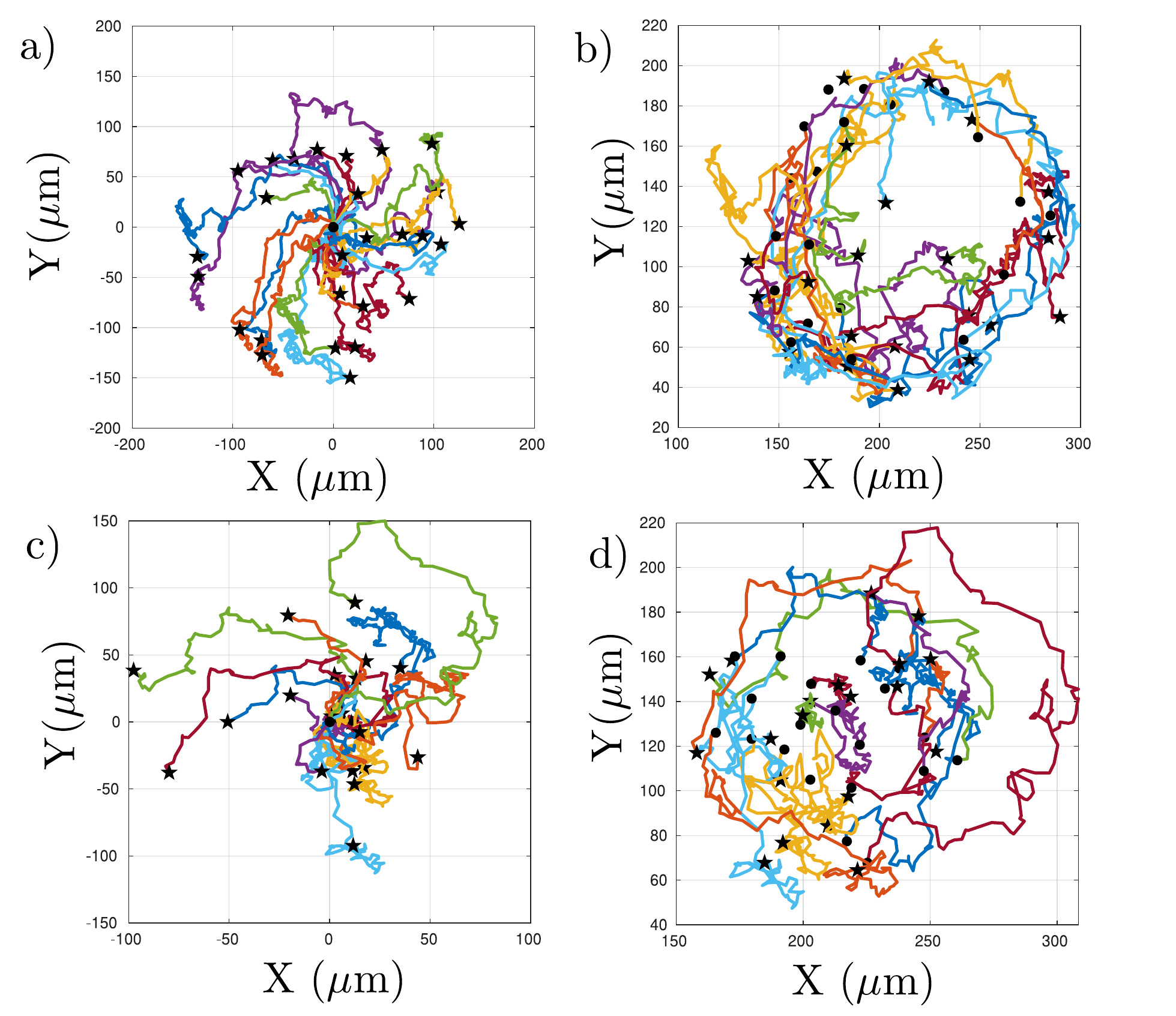}
    \caption{Trajectories of the cells moving in the outer (a, b) and inner (c, d) region of the cluster shown at the bottom with an average radius of 84 $\mu m$ and an aspect ratio around 0.83.  Solid circles (stars) show the starting (end) points.}\label{fig:5}
\end{figure*}

\subsection{Cell monolayer formation}
In addition to the growth of single cell clusters, the evolution of the T84 colony is also affected by merging of cell clusters into larger aggregates, which eventually leads to a continuous cell monolayer spanning the entire plate surface. As illustrated by the images in Fig.~\ref{fig:4}a, b, and c, the dynamics of monolayer formation was investigated by calculating, at each time step, a confluence parameter as the ratio ($A_s$) between the area occupied by the cells in the image (red regions) and the size of the whole image. The images correspond to t=0 (a), t=80 h (b) and t=160 h (c), for one of the cell samples. In Fig.~\ref{fig:4}d, $A_s$ is plotted as a function of time for several cell cultures with different initial cell density, corresponding to different values of $A_s$ at time 0. Each curve follows an exponential trend, up to almost $100\%$ confluence ($A_s=1$).

\begin{figure*}[t] % only one legend, bigger tick markers, make each graph higher
    \centering
    \includegraphics[width=0.99\textwidth]{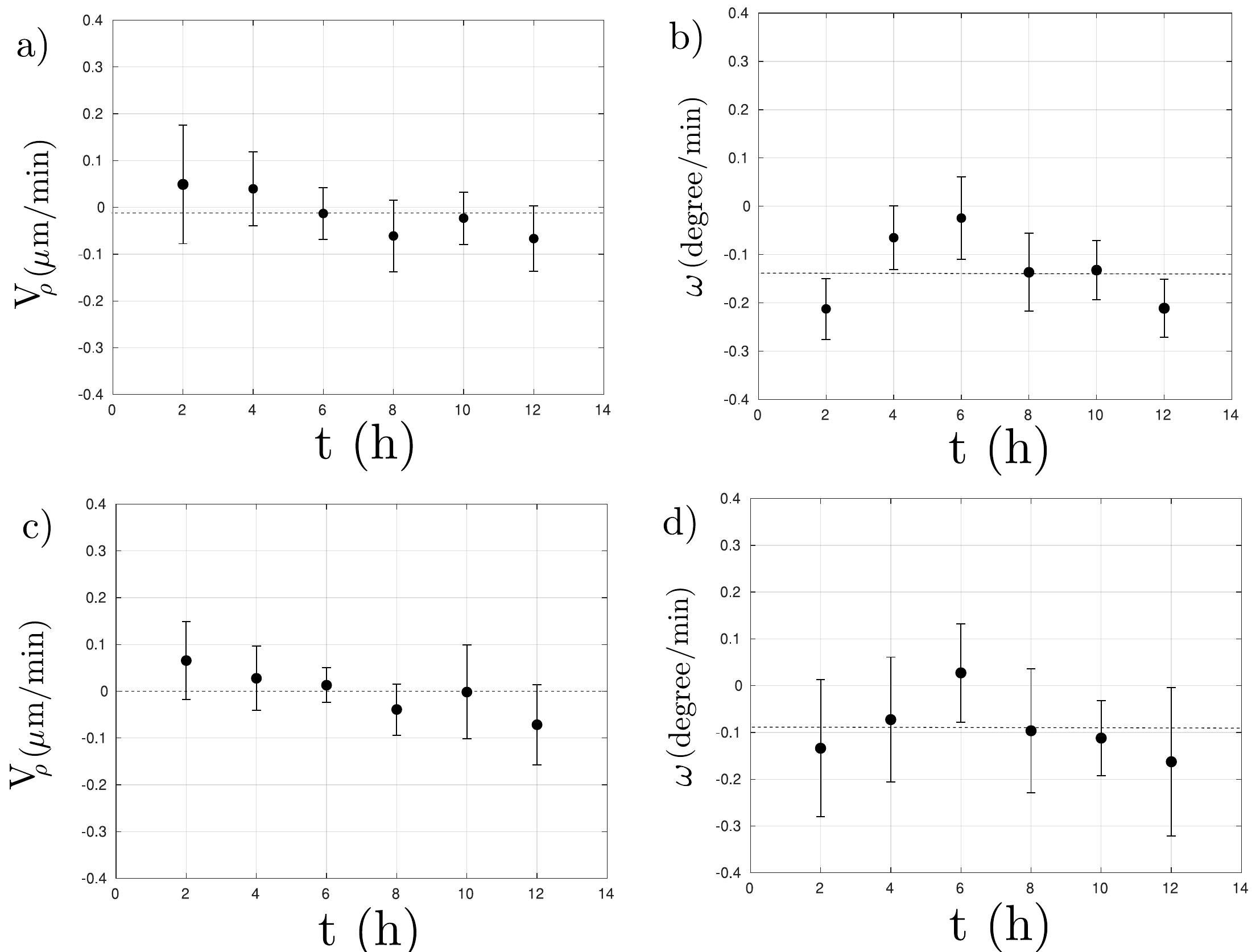}
    \caption{
Evolution in time of the radial ($V_{\rho}$), and angular ($\omega$) velocity of the cells moving in the outer (a, b) and inner (c, d) region of a cluster with an average radius of 84 $\mu m$ and an aspect ratio around 0.83. The standard deviation is reported as the error bar. }\label{fig:6}
\end{figure*}

The curves in Fig.~\ref{fig:4}d share a common increasing trend as a function of time but appear as unrelated to each other because each cell culture starts from a different initial condition in terms of cell density. This difference can be taken into account by shifting the curves horizontally to find a possible superposition of the data onto a single master curve by matching the $A_s$ values. As shown by the plot of Fig.~\ref{fig:4}e, the horizontal time shift leads indeed to the data collapse onto a master curve, thus showing that cells plated out with different initial density (different $A_s$) grow according to the same law, and the velocity of the process depends only on the actual value of $A_s$. The master curve can be fitted by the following exponential function: 

\begin{equation}
    A_s(t)= A_{s0} 2^{t/\tau},
\end{equation}
where $t$ is the time rescaled according to the horizontal shifting, $A_s(t)$ is the confluence at time t, $A_{s0}$ is the initial confluence parameter at time $t=0$, and $\tau$ is the doubling time of cell occupancy. Data from the master curve were fit according to equation (2) with $\tau$ as the only adjustable parameter (continuous line in Fig.~\ref{fig:4}).
The so obtained value of $\tau$ was found to be 50.8 h, which is within the standard error of the characteristic time of the growth of single cell clusters ($71 h \pm 19 h$, see Fig.~\ref{fig:1}). Once again, it can be noticed that the simple exponential trend is different from the sigmoidal shape of a classic logistic growth due to the limited time span of the experiment, which does not encompass the eventual levelling off of the data.

\subsection{Cell motility}
We investigated the movement of individual cells inside a cluster as a function of their position. The clusters were divided into concentric regions to evaluate possible differences in the movement of the cells in the inner and outer region of the same colony. As an example, in Fig.~\ref{fig:5} we tracked the positions of the cells moving in an almost circular cluster with an average radius of 84 $\mu m$ and an aspect ratio (minor axis/major axis ratio) about 0.83 (see image in Fig.~\ref{fig:5}e). The cluster was divided into two regions, the inner one containing all the cells within 46 $\mu m$ from the cluster center, and the external one containing the peripheral cells. In Fig.~\ref{fig:5}, the paths of the cells moving in the outer (a and b) and inner (c and d) regions are plotted in a frame of reference with the origin coincident with the initial cell position (Fig.~\ref{fig:5}a and Fig.~\ref{fig:5}c) or in the laboratory frame (Fig.~\ref{fig:5}b and Fig.~\ref{fig:5}d). The trajectories described by the cells in both regions show a circular shape, suggesting that the cells generate a coordinated collective rotation within the cluster.

Cell motion was further characterized by calculating the radial ($V_{\rho}$), and angular ($\omega$) velocity of the cells moving in the two regions over a time period of 12 h. These  are plotted as a function of time in Fig.~\ref{fig:6}a and Fig.~\ref{fig:6}b, respectively, for the cells in the outer region, and in Fig.~\ref{fig:6}c and Fig.~\ref{fig:6}d, for the cells in the core of the cluster.

\begin{figure*}[t] % only one legend, bigger tick markers, make each graph higher
    \centering
    \includegraphics[width=0.99\textwidth]{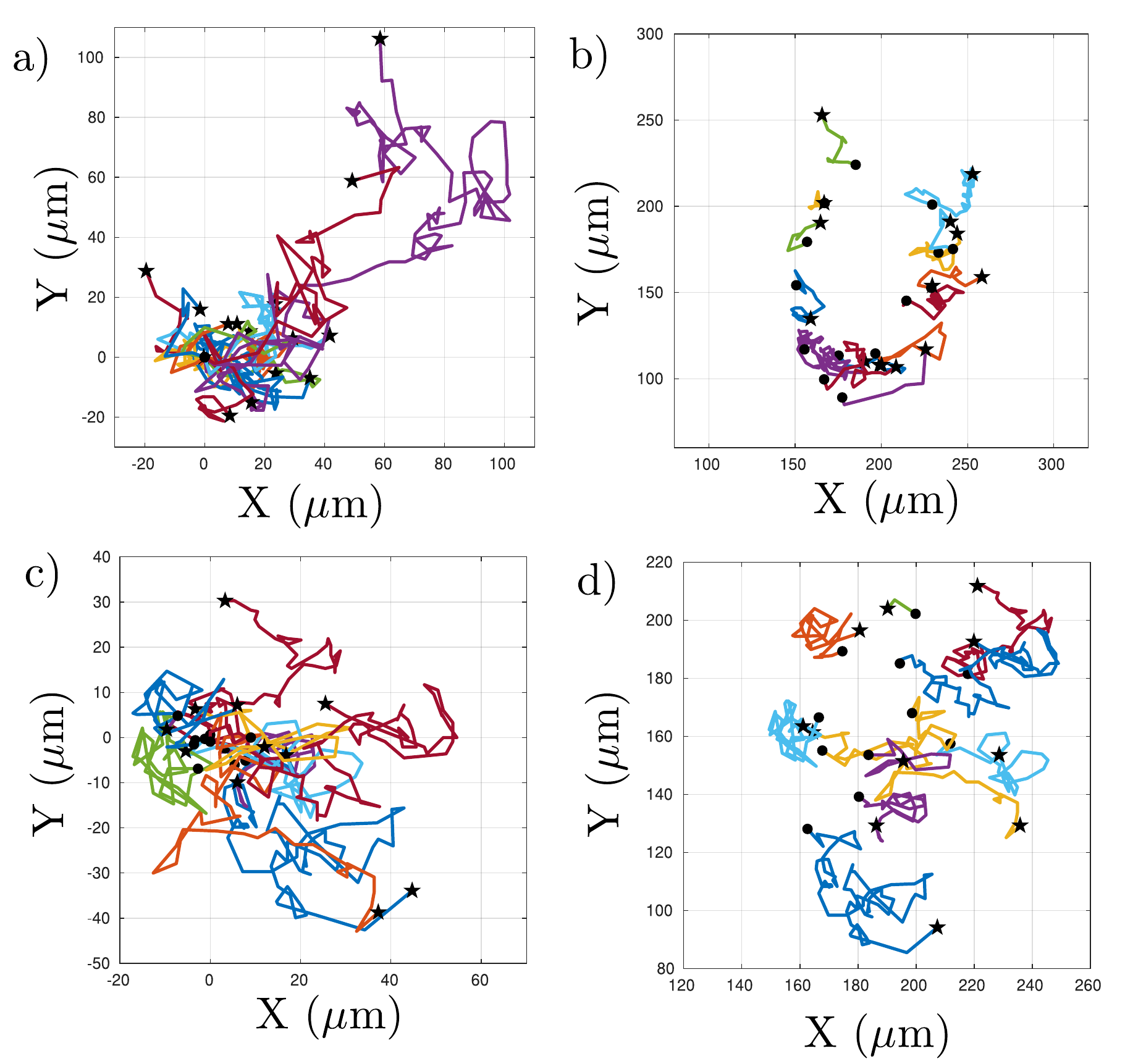}
    \caption{
Trajectories of the cells moving in the outer (a, b) and inner (c, d) 
region of the cluster shown at the bottom with an average radius of 70 $\mu m$ and an aspect ratio around 0.65. 
Solid circles (stars) show the starting (end) points.}\label{fig:7}
\end{figure*}

Our data suggest that the movement along the radial direction of the cells in the outer as well as in the inner region of the colony is quite limited, which is consistent with a free expansion regime. Concerning the radial velocity, by looking at different clusters an average value of 0.04 $\mu m/min$ can be obtained (which was used above to estimate the transition between the free growth and the proliferation inhibition regimes). Hence, the cells migrating along the edge of the aggregate do not invade the core region, and vice versa. 

The cells moving in the outer region of the cluster show higher tangential velocity compared to the cells in the core, and display a higher coordination in the rotatory movement as suggested by the lower amplitude of the error bars. The time evolution of the angular velocity of the cells on the edge and in the core region of the cluster exhibits the same trend, suggesting that the outer cells drag the inner ones, and coordinate their rotatory movement. Overall, the cells of the entire cluster move in a concerted fashion like a rotating disk. The coordinated rotational movement of the cells in colonies with a round shape has been already shown in previous works \cite{doxzen2013guidance,malet2015collective}, but only for cells confined in micropatterns. To the best of our knowledge, this is the first report of a systematic study of CAM in freely expanding cell colonies. 
Since clusters exhibit different shapes, which can have an effect on coordinated cell rotation, we repeated the tracking analysis of cell positions in a colony with an 
average radius of 70 $\mu m$ and an aspect ratio $0.65$, i.e., with a less circular shape. In this case too, the cluster was divided into two regions, the inner one being a circle with a radius of 38 $\mu m$. In Fig.~\ref{fig:7}, the trajectories of the cells moving in the outer (a and b) and inner (c and d) region are plotted relative to a common origin (Fig.~\ref{fig:7}a and Fig.~\ref{fig:7}c) or to their actual initial position (Figure Fig.~\ref{fig:7}b and Fig.~\ref{fig:7}d). The plots in Fig.~\ref{fig:7} show that the cell trajectories are less coherent with respect to the ones of a more circular cluster (see Fig.~\ref{fig:7}).

\begin{figure*}[t] % only one legend, bigger tick markers, make each graph higher
    \centering
    \includegraphics[width=0.99\textwidth]{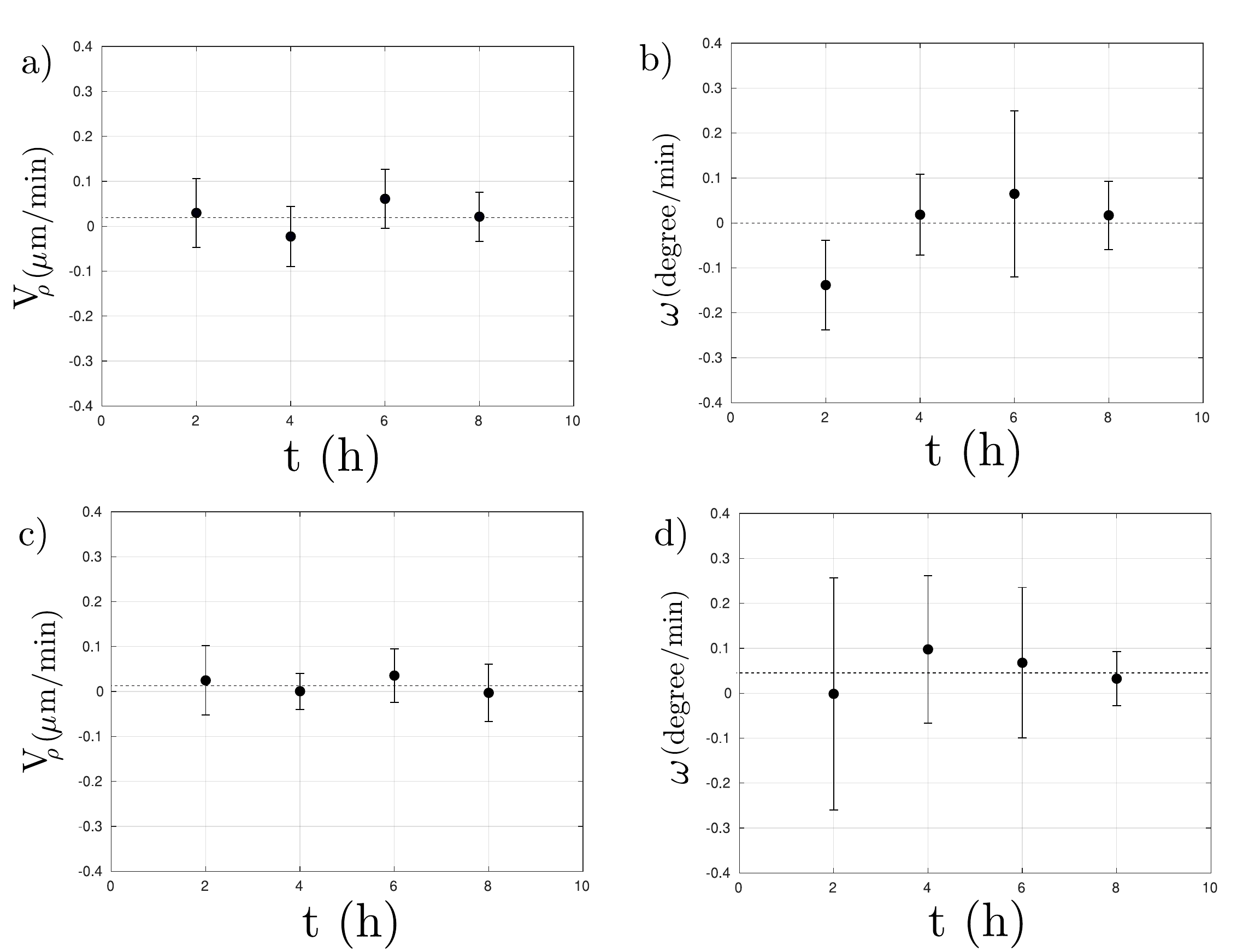}
    \caption{Evolution in time of the radial ($V_{\rho}$), and angular ($\omega$) velocity of the cells moving in the outer (a, b) and inner (c, d) region of a cluster with an average radius of 70 $\mu m$ and an aspect ratio around 0.65. The standard deviation is reported as error bar.}\label{fig:8}
\end{figure*}

Such a result is confirmed by the radial ($V_{\rho}$), and angular ($\omega$) velocity of the cells moving in the outer (Fig.~\ref{fig:8}a and b) and inner (Fig.~\ref{fig:8}c and d) regions of the cluster. The angular velocities of both the outer and the inner region have a time average quite close to 0, thus showing that a coordinated rotation is very small, if any at all. Furthermore, the error bars of the angular velocities (Fig.~\ref{fig:8}c and Fig.~\ref{fig:8}d) are larger than the ones of the more circular cluster of Fig.~\ref{fig:5} and Fig.~\ref{fig:6}. This result shows a more random motility in the less circular cluster. 

\begin{figure*}[t] % only one legend, bigger tick markers, make each graph higher
    \centering
    \includegraphics[width=0.99\textwidth]{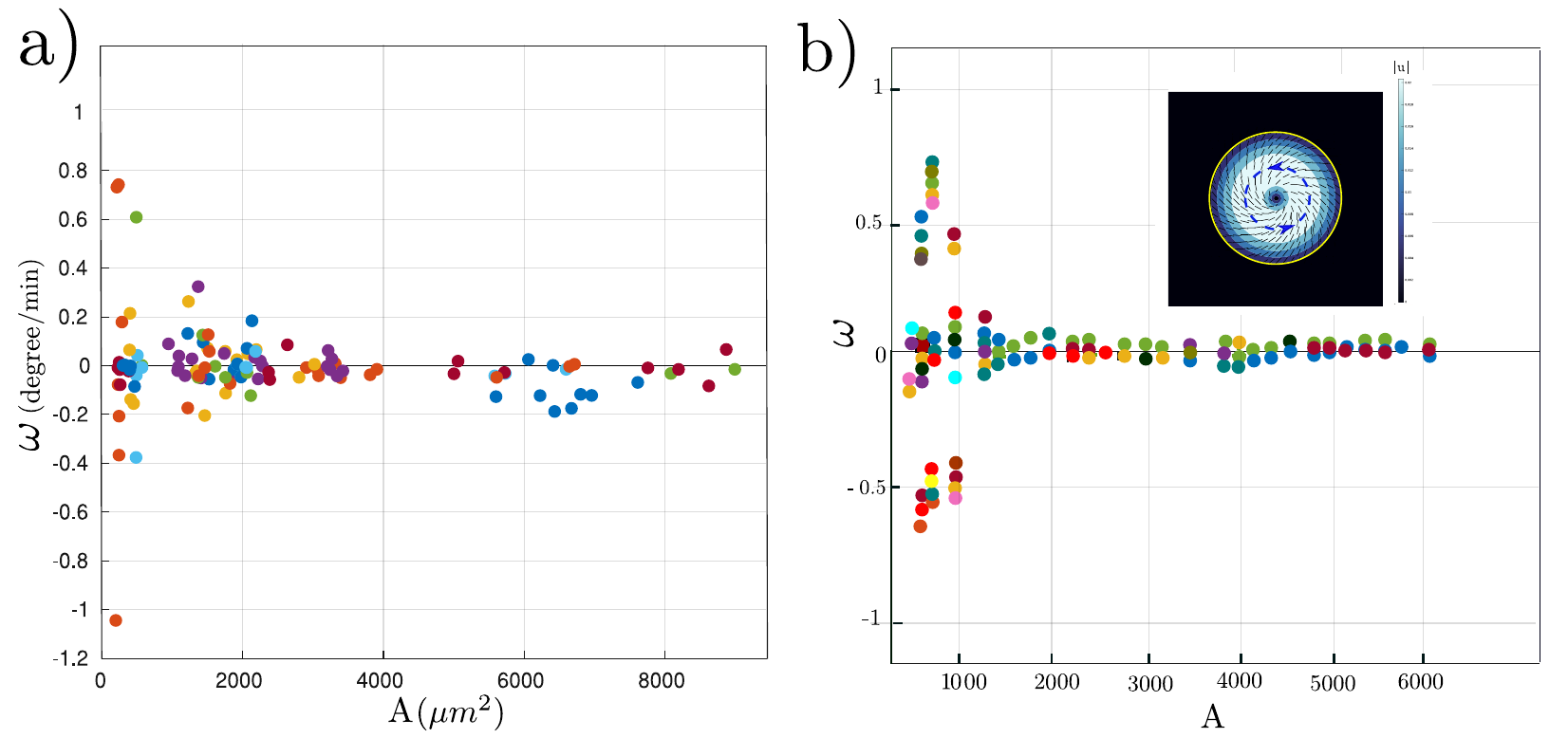}
    \caption{Angular cell velocity as a function of cluster area in experiments (a), and simulations (b). 
    To compare the rotational velocity with the
experimental data, we matched the initial area and angular velocity in simulations with the initial area and angular velocity in the experiment and rescaled the axes accordingly. In both cases, colonies show collective rotation for intermediate values of the droplet area. Inset in part (b) shows a rotating colony in simulations. Color shows the magnitude of the velocity field, blue dashed arrows show the direction of the rotation, and yellow shows the interface of the colony.}\label{fig:10}
\end{figure*}

It is tempting to conjecture that boundary curvature variations in a more irregular (non-circular) cluster lead to non-uniform motile behavior of the cells. This finding, combined with a lack of radial motion independently of cluster shape (which shows that cells tend to stay within a defined radial region inside each cluster), can be explained by the effect of cell-cell junctions, which hold them together, thus generating a solid-body rotation. Irregular cluster shapes are mostly generated by the merging of different clusters when they get in contact due to their growth. However,  no systematic trend of aspect ratio vs cluster size was found (data not shown for the sake of brevity). 
The effect of cluster size on rotation is illustrated in Fig.~\ref{fig:10}, where the average angular cell velocity in a cluster is plotted as a function of cluster area for 30 cell clusters from two independent experiments.

The continuous line is a linear fit of the data and is coincident with the horizontal axis, thus showing that the average angular cell velocity among all the clusters is zero. However, the absolute value of angular velocity is a decreasing function of cluster size and becomes essentially zero at a value of cluster area of about $3000 \mu m^2$, which corresponds to a diameter of $60 \mu m$. This dependence of  on $A_{cluster}$  is in qualitative agreement with the results from the literature on MDCK cells in micropatterned substrata \cite{doxzen2013guidance}, where the collective rotation was found to be persistent at higher cell densities, but with a decrease of the average cell velocity with cluster size. Interestingly, in this previous study the synchronized collective rotation was only observed in 100 and 200 $\mu m$ diameter regions (close to the correlation length found in unconfined MDCK epithelial sheets \cite{poujade2007collective}), while transient vortices reminiscent of active turbulence were found in larger regions. The lower value of the critical size for coordinated cell rotation observed in our experiments (60 vs 200 $\mu m$) can be due to the different cell line (T84 vs MDCK) and/or to the free vs confined expansion regime.
Another interesting feature of the data in Fig. \ref{fig:10} is that clockwise (negative) and counterclockwise (positive) rotations are equally likely, since the average angular velocity is zero. This is not a trivial result, since some asymmetry in collective rotation, with a switch in CAM direction from time to time, has been observed in confluent MDCK cells elsewhere \cite{deforet2014emergence} and attributed to cell chirality, which is a phenotype-specific property assuming a clockwise value for several (but not all) cell types from fibroblast to endothelial cells \cite{wan2011micropatterned}. Our data do not support a special orientation in cell rotation for the T84 cell line under investigation.

%\section{Simulation Methods}\label{secequations}

\section{Simulation Results}
We next checked whether the cell rotation could be described in terms of the theories of active nematics. We used the continuum active nematohydrodynamic equations of motion (see Methods) to consider growing, initially isotropic cell colonies. We found that, once the size of the colony has reached a threshold value, active flows induced nematic order and the colonies could start rotating (see inset in Fig.~\ref{fig:10}b). The angular velocity for several realisations of the simulations is shown in Fig.~\ref{fig:10}b and can be compared with the experimental results in Fig.~\ref{fig:10}a. In agreement with the experiments, we found that  smaller colonies show collective rotation and the angular velocity decreases with colony area and is zero for larger colonies.  The figure also shows that the direction of the rotation is randomly selected between clockwise and anti-clockwise, and does not change over time, in agreement with the experiments, and as expected as  the active stress does not have any chirality.

\section{Conclusions}
In this work, we show for the first time that coordinated angular motion is a feature of freely expanding epithelial cells and does not require spatial confinement of the cells on micropatterned substrata. One of the main implications of this result is that physical confinement of cells at the boundary of a cluster is not a necessary condition for the development of CAM. This does not rule out a possible role of boundary cells to initiate a coherent rotation which propagates to the inner parts of a cluster. A further argument supporting a special status of boundary cells is their morphology (more flattened and with several protrusions exploring the surrounding environment), which is quite different with respect to the one exhibited by cells inside a cluster. However, a special role played by boundary cells in eliciting cluster rotation is questionable. We show indeed that the observed experimental trends are in excellent agreement with predictions based on active nematic theories. In particular, we found, both by experiments and simulations, that the angular velocity was a decreasing function of cluster size. The spontaneous tendency of epithelial cells to rotate in a synchronized fashion could reflect a morphogenetic mechanism, such as the formation of acini in the mammary glands and the tangential cell orientation found on the wall of intestinal microvilli. More work is needed to unravel these mechanisms and elucidate the drivers of coordinated cell rotation at the molecular scale.

\end{multicols}

\enlargethispage{20pt}

%\ethics{Insert ethics text here.}

\dataccess{This article has no additional data.}

\aucontribute{Insert author contribute text here.}

\competing{We declare we have no competing interest.}

\funding{No funding has been received for this article.}

\ack{The insightful contribution of Luigi Maiuri to the planning and interpretation of the experimental results and for critical reading of an earlier version of the manuscript is deeply acknowledged. The undergraduate students Roberta Liuzzi, Federica Borgia, Marilena Cerbone, Maria Giuseppa Coppola, Federica Granata contributed to the analysis of the experimental data during their bachelor thesis. M. R. N. acknowledges the support of the
Clarendon Fund Scholarships.}

\disclaimer{We declare we have no competing interest.}

%%%%%%%%%% Insert bibliography here %%%%%%%%%%%%%%

\printbibliography

\end{document}